\documentclass[12pt]{article}
\usepackage[dvips]{graphicx}
\title{
Spin-orbit and exchange interaction in surface quantum wells on gapless
semimagnetic semiconductor HgMnTe
}
\author{
V.F. Radantsev  \thanks{e-mail: victor.radantsev@usu.ru}\\
\small\em Institute of Physics and Applied Mathematics, \\
\small\em Ural State University, Ekaterinburg 620083, Russia\\[10pt]
\large
A.M. Yafyasov, V.B. Bogevolnov and I.M. Ivankiv\\
\small\em Department of Solid State Electronics, Institute of Physics, \\
\small\em St. Petersburg State University, St. Petersburg 198904, Russia
}
\date{\today }

\begin{document}

\maketitle

\begin{abstract}
The first study of two-dimensional electron gas in surface layers on HgMnTe
with inverted bands is carried out experimentally and theoretically. It is
shown that the structure of investigated capacitance magnetooscillations in
HgMnTe MOS structures is fully similar to the one in the non-magnetic
narrow-gap semiconductor HgCdTe and the sole effect due to exchange
interaction is the temperature shift of beat nodes. The information about
exchange effects is obtained only due to our modeling of oscillations,
because any pronounced changes in the position of oscillations are not
observed and the separate spin components are not resolved. The Landau
levels are calculated in the framework of concept we developed previously
for the description of subband dispersions in zero magnetic field $B=0$. The
new parameters (like those of $T/\sqrt{eB}$) arise in the theory of
magnetooscillations in the semiconductors with quasirelativistic spectrum in
contrast to the case of parabolic bands. The modeling shows that the
spin-orbit splitting far exceeds a contribution due to exchange interaction.
The calculated amplitudes of ``partial'' oscillations for different spin
branches of spectrum are essentially different in accordance with the
observed difference in the intensity of corresponding lines in Fourier
spectra. The comparison between experiment and theory for different
temperatures and parameters of exchange interaction is reported. The
dominant mechanisms of the scattering responsible for the broadening of
Landau levels are discussed.
\end{abstract}

\medskip

{\bf PACS numbers:} 73.20.Dx, 73.40.Qv, 71.70.Ej

\section{Introduction}

\label{sec:intro}

There are two aspects that cause the peculiar features of, and the interest
in, the two-dimensional (2D) electron gas in narrow-gap diluted magnetic
(semimagnetic) semiconductors (DMS). One stems from the $s,p-d$ exchange
interaction between band electrons and localized magnetic moments \cite{Furd}.
This interaction results in changed spin splitting of the band states, which
can be varied by such external factors as a magnetic field and temperature.
The other is due to the peculiarities inherent to Kane semiconductors with
small gap, leading to such relativisticlike effects as non-parabolicity,
kinetic confinement (motional binding \cite{Doezema}), spin-orbit (SO)
splitting and resonant interband mixing by surface electric field
\cite{Takada,Rad96,German}.
What is important in the context of the specificity of
2D electronic systems involving DMS, is that both exchange and SO
interaction cause the rearrangement of the spin structure of Landau levels
(LL's). Although historically the first studies of 2D electron gas in DMS
were performed for metal-insulator-semiconductor (MIS) structures based on
HgMnTe, \cite{Grab84} the available experimental results are mostly for the
grain boundaries in HgMnTe and HgCdMnTe with positive Kane gap
$E_{g}>100$\,meV at Mn content typically $x=0.02$ (for the higher
$x$, the exchange interaction exhibited itself poorly,
that was attributed to
antiferromagnetic interaction betwen Mn
ions) \cite{Furd,Grab84a,Grab89,Grab93}.
This is due to low electron mobility in
previously investigated MIS structures. At the same time, the inversion
layers in MIS structures are of particular experimental interest because of
the possibility of controlling the depth of surface quantum well by gate
voltage, and because of the relative ease and accuracy of the description of
surface potential (in the case of the bicrystals, the additional poorly
verified assumptions have to be used to describe a self-consistent potential
near grain boundaries \cite{Sobk89}). An important point is that these
results can be compared with the data for MIS structures based on narrow-gap
HgCdTe \cite{Rad96,Rad85}, being non-magnetic analogue of narrow-gap DMS.

As regards their theoretical description, the subband calculations were
carried out only for DMS with direct but not inverted bands and without
allowance for spinorlike effects \cite{Furd,Grab84}. However, the SO
splitting in asymmetrical quantum wells at zero magnetic field (this
phenomenon in itself is of much current interest
\cite{Rad96,German,Rashba,Rad89,Luo,Das,Engels,Nitta,Heida,Wissing}) leads to the
rearrangement of subband magnetic levels. In narrow-gap semiconductors, the
perturbation of magnetic spectrum is so drastic that SO interaction cannot
be neglected in the theoretical treatment. It must be stressed that SO
splitting, as we shall see in Sec.\,III, exceeds by far exchange interaction
contribution so it cannot be considered as a correction to the exchange
interaction. It is clear also that a treatment based on the quasiclassical
quantization in a magnetic field of subband spectrum (calculated at $B=0)$
is unsuitable for the description of exchange interaction effects. A more
rigorous theoretical consideration of the LL's structure is required.

In this paper, the peculiarities of 2D electron gas due to exchange and SO
interaction are studied in inversion layers on Hg$_{1-x}$Mn$_{x}$Te.
with a small Mn content. At $x<0.08$ HgMnTe has inverted bands (i.e. becomes
semimetal) and traditional galvanometric methods cannot be used because of
the shunting of surface conductance by the bulk. We employed the
magnetocapacitance spectroscopy method, which is applicable to
semiconductors with any sign of the Kane gap. The parameters of the samples
and the experimental data relating to the capacitance oscillations versus
gate voltage and magnetic field and their temperature evolution are
presented in Sec.\,II. In Sec.\,III, we present the theoretical model. The
treatment of LL's in 2D subbands is based on the further development of the
concept we offered previously for the description of subband spectrum
at $B=0$. The density of states (DOS) in a magnetic field is described
neglecting the mixing between LL's and assuming a Gaussian shape of each
level. In this section an analytical expression for oscillations of the
differential capacitance of space charge region in the low-temperature range
is also obtained using WKB approach. In Sec. IV, the results of the computer
modeling of capacitance oscillations are presented. The results of a
comparison of the experimental data and theoretical calculations for
different temperatures and parameters of exchange interaction are discussed.
The parameters of\ the broadening of LL's are determined from a fitting of
the amplitudes of calculated oscillations to their experimental values. The
dominant mechanisms of the scattering responsible for broadening of LL's are
discussed.

\section{Samples and experimental results}

\subsection{Samples and experimental methods}

In this work the inversion layers in MIS structures fabricated from
$p$-Hg$_{1-x}$Mn$_{x}$Te single crystals were investigated.
No impurities were
introduced intentionally and no post growth annealing was performed.
Deviations from stoichiometry, and thus the type of the conductivity, were
controlled by mercury partial pressure during the growth process. The
Hall-effect measurements were performed at variable temperatures and
magnetic field strengths. After removing the Hall and tunnel contacts (see
below), the substrates were mechanically polished and etched in a
0.5\,bromine-methanol solution. Several methods such as anodic
oxide formation,
silicon oxide and Al$_{2}$O$_{3}$ deposition, and the Langmuir-Blodgett film
technique have been used for forming an insulating film in MIS structures.
The gate electrodes of the typical area $\sim $ $5\times 10^{-4}$ cm$^{2}$
were formed evaporating Pb. The differential capacitance $C$ and
derivative $dC/dV_{g}$ on gate voltage $V_{g}$ of the capacitors
were measured in the dark, typically at 1\,MHz and a test signal
amplitude of 5\,mV.

The capacitance magnetooscillations due to the magnetic quantization of 2D
electron gas were observed in all the above HgMnTe MIS structures. It was
shown that the general shapes of the oscillations at the same carrier
surface density and Mn content are similar. In the following we present the
results for the structures with a $\sim 80$ nm thick anodic oxide film,
grown in a solution of 0.1M\,KOH in 90\% ethylene glycol / 10\% H$_{2}$O
at $0.1$\,mA$\cdot$cm$^{-2}$.
There are several reasons for such a choice: (i) the
amplitudes of oscillations in these structures are the highest owing to the
large value of insulator capacitance (this is caused by the large value of
dielectric constant of the anodic oxide), (ii) the highest surface carrier
densities are achieved at low gate voltages $V_{g}=10-15$\,V, and (iii) the
dielectric constant of oxide is close to that of a semiconductor so the
contribution of image forces in surface potential can be neglected in the
calculations.

We investigated samples with different Mn content ($x=0.024,0.040,0.060$ and
$0.1$). Kane gap $E_{g}$ and Kane effective mass $m_{b}$ (and therefore $x$)
were determined independently by the tunnel spectroscopy method for a
comparison of band parameters in the bulk with those in the vicinity of the
surface. The discrepancy is within the accuracy of the analysis ($\Delta
x\sim 0.002\div 0.003$). Because the tunnel contacts and studied MIS
capacitors were produced using identical technology and differ only by
thickness of insulator (Langmuir-Blodgett film or an oxide) this agreement
testifies that the surface layers are chemically close to the bulk. The
similarity of the results for the structures with different insulators (with
different fabrication methods) supports this conclusion. The fact that in
the small surface concentration range the measured cyclotron masses in 2D
subbands extrapolate to the bulk value $m_{b}$ is direct evidence of an
absence of noticeable decomposition in 2D layer during the structure
fabrication process.

In the following we shall restrict our consideration to the results for
HgMnTe with $x\approx 0.04$ ($Eg=-100\pm 5$\,meV). The amplitudes of the
oscillations for other samples are much less even at 4.2\,K and rapidly
decrease with increasing temperature. In the case of $x\approx 0.024$ this
is caused by the small cyclotron energy due to a large Kane gap. In the
cases $x\approx 0.06$ and $x\approx 0.1$ it is due to the large doping level
of available materials. As a result, we could not obtain reasonably accurate
information about the oscillation's temperature evolution, in which the
specificity of DMS is manifested. As to the measurements at $T=4.2$\,K, the
subband parameters extracted from oscillations for these samples are similar
to those for HgCdTe with the same band parameters and agree well with the
theory. On the other hand, the samples with $x=0.04$ are best suited to the
purpose of this first study aimed at investigating the peculiarities of 2D
electron gas in DMS with inverted bands, in which (i) the effects of SO and
exchange interaction are expected to be more clearly pronounced and (ii) the
results can be compared with those for well studied surface layers on
gapless HgCdTe with $E_{g}\sim -(50\div 100)$\,meV \cite{Rad96,Rad93}. For
small gap $\left| E_{g}\right| <100$\,meV the parameters of 2D subbands
depend only weakly on $E_{g}$\ (except in the case of small subband
occupancies).\cite{Rad96,Rad93} By contrast, the subband parameters are more
sensitive to the doping level. For this reason, we present the results for
two samples with $N_{A}-N_{D}=1.2\times 10^{16}$\,cm$^{-3}$ (sample $S1$) and
$N_{A}-N_{D}=1.5\times 10^{17}$ $cm^{-3}$ (sample $S2$). The Hall mobility
of holes is of the order of 2500\,cm$^{2}$/Vs at 4.2\,K.
The experimental data
are presented below for capacitors with the gate area
$S=7.2\times 10^{-4}$\,cm$^{2}$
and insulator capacitance $C_{ox}=136.5$\,pF for sample $S1$,
and with $S=7.7\times 10^{-4}$ and $C_{ox}=155.1$\,pF for sample $S2$.

\subsection{Capacitance measurements in perpendicular magnetic fields}

\begin{figure}
\centerline{\includegraphics[width=10cm]{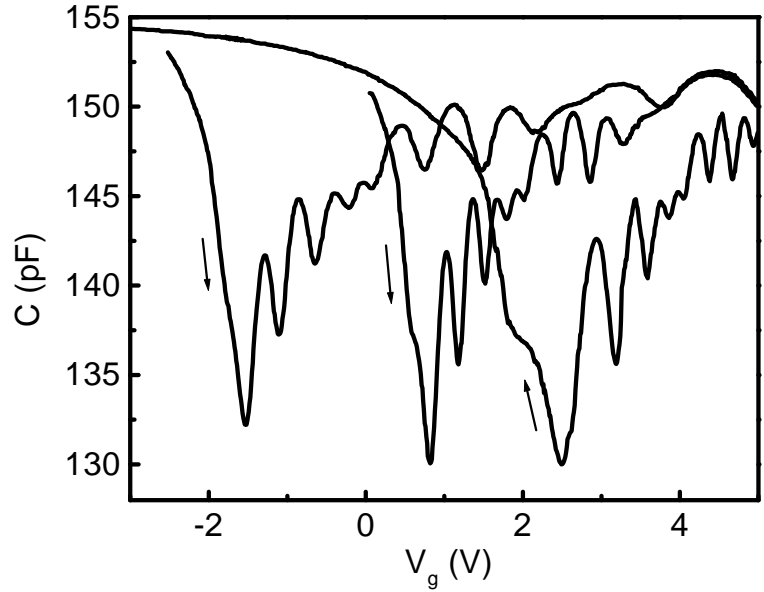}}
\caption{Capacitance-voltage dependencies in magnetic field $B=4.5T$
perpendicular to 2D plane for sample $S2$ at different gate voltage sweep.
The arrows indicate the sweep direction. The sweep rate is 2\,V/Min.}
\label{fig1}
\end{figure}

Fig.\,\ref{fig1} shows the capacitance-voltage characteristics at $T=4.2$K
in magnetic field $B=4.5$\,T perpendicular to the 2D layer
for the sample $S2$. $C(V_{g})$ characteristics are those
of typical low-frequency behavior.
This is due to the absence of a gap between conduction and valence (heavy
hole) band in gapless semiconductors. As a result, 2D electrons in
inversions layers are in equilibrium with the ac ripple and contribute
predominantly to the measured capacitance under inversion band bending. The
low-frequency conditions with respect to the minority carriers are satisfied
in all the investigated frequency range 30\,kHz$\div $5\,MHz. The wide
hysteresis loop and the dependence of $C(V_{g})$ characteristics on the rate
of voltage sweep are observed. The capacitance changes in time because of
flat-band voltage shift $\Delta V_{fb}$, which is close to logarithmic in
time. The time constant is of the order of a few minutes and is almost
independent of the temperature. The hysteresis effects point to charge
tunnel exchange between the semiconductor and the slow
traps in insulator \cite{Adar}.
A history dependence and instability are manifested in all the
investigated HgMnTe-based MIS structures. Such behavior is contrary to that
of HgCdTe and HgTe-based structures with the same insulators.

The voltage dependence of the charge density $eN_{s}(V_{g})$ induced in
inversion layer is sublinear. This is well demonstrated by the
non-equidistance of quantum oscillations of the capacitance $C(V_{g})$ (see
Fig.\,\ref{fig1}). The tunneling of electrons from the 2D layer into oxide
causes a saturation of $N_{s}(V_{g})$ dependence at $V_{g}-V_{fb}$ $\approx
(10\div 15)$\,V.
As a result, the $N_{s}$ range accessible for investigations
is limited by the value $(3\div 4)\times 10^{12}$ cm$^{-2}$ (in HgCdTe,
values of $N_{s}$ up to 10$^{13}$ cm$^{-2}$ can be obtained). Although the
hysteresis effects hamper the measurements, the discussed physical results
are not affected by the instability of band bending. Such instability is
caused by the transient processes but not by degradation. In order to assure
the stability of band bending during the measurement of $C(B)$ oscillations,
the sample was held at given capacitance (or voltage) for 5-15\, minutes. The
identity of $C(B)$ plots registered at increasing and decreasing magnetic
field (i.e. at different times) was examined for each $C(B)$ curve. When the
temperature (or angle) dependencies of $C(B)$ oscillations were measured,
the long term stability was checked by the repetitive measurement of initial
(for given measurement cycle) $C(B)$ plots (see Fig.\,\ref{fig5}). Although
the $C(V_{g})$ and $dC/dV_{g}(V_{g})$ characteristics are history dependent,
they are completely repeatable, if the voltage range, rate and direction of
the sweep are the same (see Fig. \,\ref{fig6}).

As can be seen in Fig.\,\ref{fig1}, the magnitudes of capacitance in the $%
C(V_{g})$ oscillation extrema, corresponding to the same LL's number, are
the same for the curves with different $V_{fb}$. $C(B)$ oscillations (and
consequently subband occupancy and surface potential) measured at the same
magnitude of capacitance in a zero magnetic field $C(0)$ are also identical
no matter what the voltage (the value of the latter for any given $C(0)$ is
determined by the flat-band voltage, which is a history- and
time-dependent). When the dc gate voltage (or flat-band voltage at the same $%
V_{g}$) is changed, the filling of interface states is also changed but does
not respond to the ac ripple, i.e., the interface states do not contribute
to the capacitance. This takes place for all frequencies and temperatures
and testifies that the high-frequency conditions with respect to interface
states are satisfied \cite{Adar,Nicoll}. Thus there is ``one-to-one
correspondence'' between $C(0)$, band bending and surface density of 2D
electrons $N_{s}=\sum $ $N_{i}$ ($i$ is the 2D subband number). Owing to
this we can reproduce $C(B)$ curves (characterized by the $C(0)$ values) at
all times. The subband parameters are presented below as functions
of $N_{s}$. Contrary to dependencies vs. $V_{g}$, these dependencies
are not affected
by the hysteresis effects or any specific parameters of MOS capacitors and
are common to the given HgMnTe sample. It may be noted that the hysteresis
has some positive points also. We have the possibility to investigate 2D
electron gas in the same surface quantum well on the same sample but with a
different interface charge. Particularly, it is important for the
investigation of scattering mechanisms.

\begin{figure}
\centerline{\includegraphics[width=10cm]{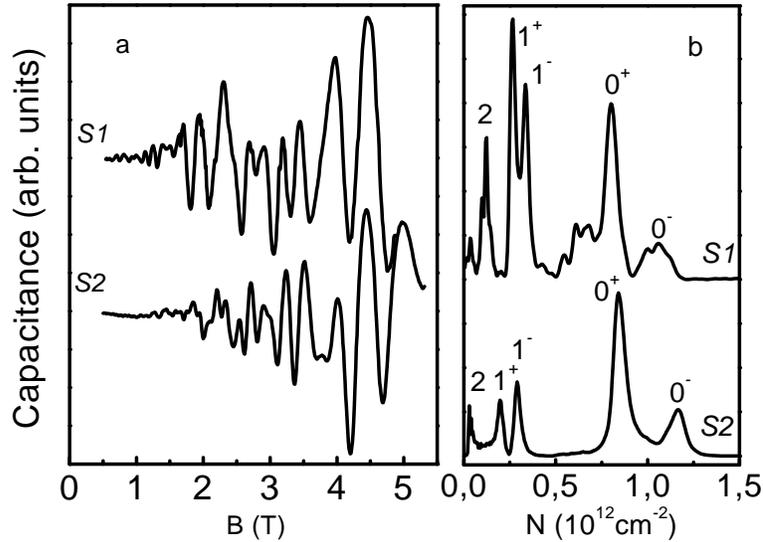}}
\caption{Capacitance oscillations ($a$) and their Fourier spectra ($b)$ for $%
S1${\ (}$N_{s}=2.76\times 10^{12}$\,cm$^{-2}$) and $S2$ {(}$N_{s}=2.56\times
10^{12}$\,cm$^{-2}$) {samples} at close surface densities $N_{s}$. The
frequency (horizontal) axis in the right panel is calibrated in the
concentration units but is not normalized to the spin degeneracy. The well
defined split of Fourier spectra into two lines corresponding to high-energy
$(i^{+})$ and low-energy $(i^{-})$ branches of subband spectra is observed
for subband levels $i=0$ and $i=1$.}
\label{fig2}
\end{figure}

\begin{figure}
\centerline{\includegraphics[width=10cm]{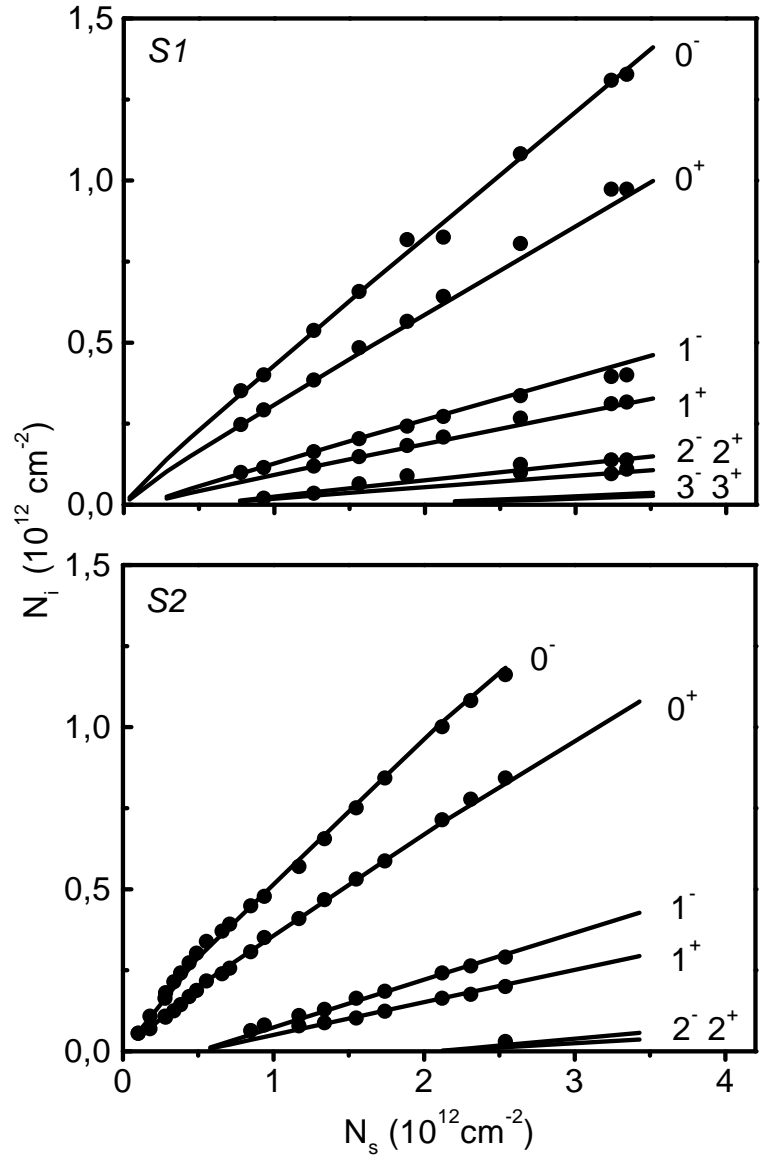}}
\caption{Calculated (lines) and measured (points) distribution of 2D
electrons among the spin-split subbands $i^{\pm }$ for $S1$ and $S2$
samples. The theoretical dependencies are calculated 
as in Ref.\,\cite{Rad96}.}
\label{fig3}
\end{figure}

\begin{figure}
\centerline{\includegraphics[width=10cm]{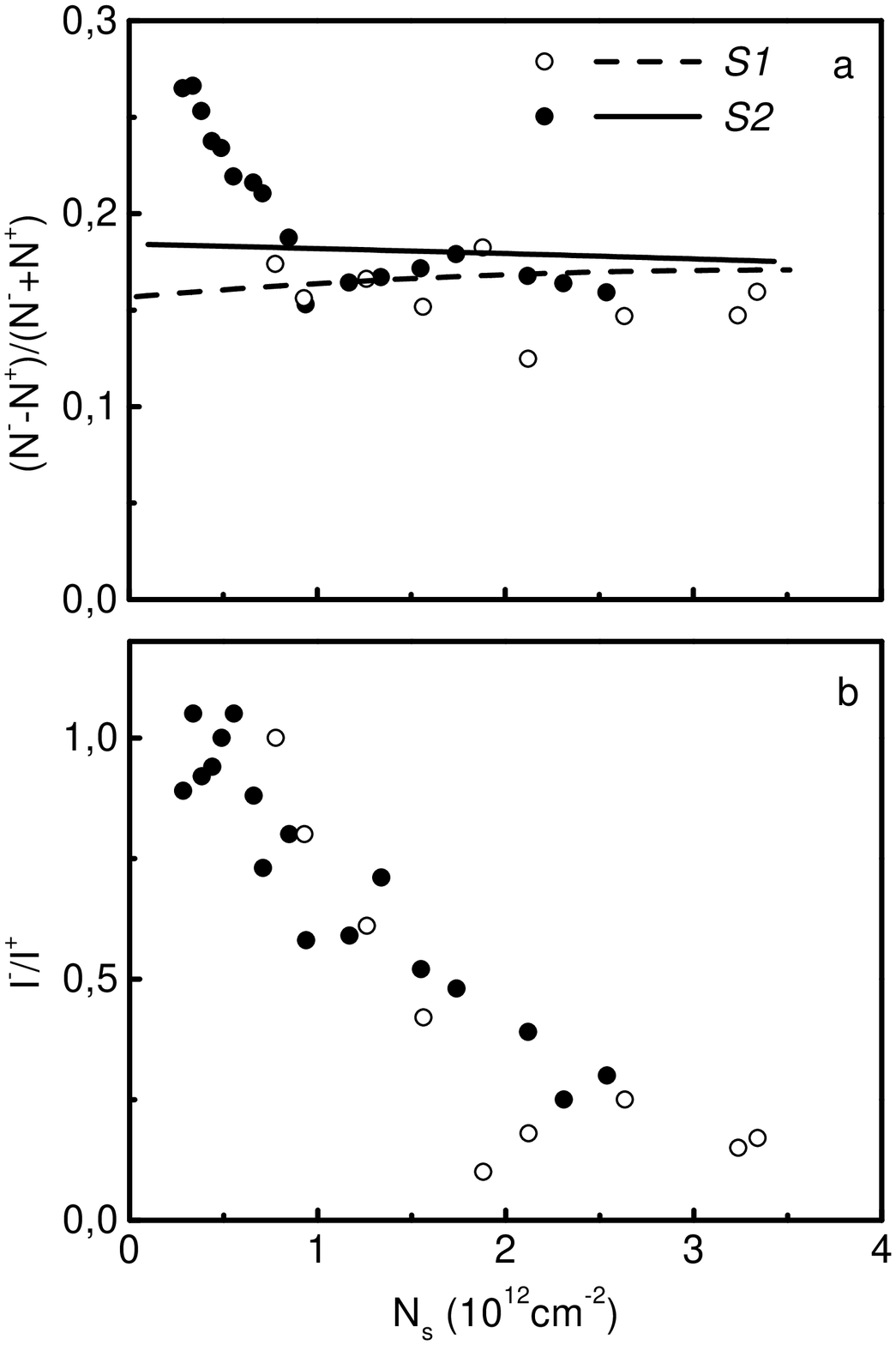}}
\caption{The relative difference of spin-split subbands occupancies (a) and
the ratio of the intensities of Fourier lines for the low-energy ($I_{i}^{-}$%
) and high-energy ($I_{i}^{+}$) branches (b) versus $N_{s}$ for ground
subband $i=0$ for $S1$ and $S2$ samples. The lines in the upper panel are
theoretical dependencies calculated as in Ref.\,\cite{Rad96}.}
\label{fig4}
\end{figure}

Typical $C(B)$ oscillations for both samples at almost equal $N_{s}$ are
presented in Fig.\,\ref{fig2} together with their $1/B$ Fourier transforms.
As in the case of gapless HgCdTe \cite{Rad96}, the individual spin components
have not been observed in the oscillations at any $N_{s}$ even for lowest
LL's. On the other hand, the oscillation beats and the Fourier spectra
demonstrate distinctly the presence of two frequencies connected with the SO
splitting of each 2D subband. The additional structure in Fig.\,\ref{fig2}
near the Fourier lines corresponding to $i=0$ subband for sample $S2$
results from mixed harmonics. This structure is suppressed if the range of
small magnetic fields (in which the oscillations relevant to the higher
subbands are dominant) is excluded from Fourier analysis. For a sample $S1$,
the intensities of $i=1$ and $i=2$ lines are small, and the above features
are practically indistinguishable in the scale of Fig.\,\ref{fig2}. 
The surface densities in the spin-split subbands $N_{i}^{+}$ and $N_{i}^{-}$
determined from Fourier transforms are plotted in Fig.\,\ref{fig3} as
functions of $N_{s}$. The carrier distribution among 2D subbands is
different for the two samples. The concentrations $N_{s}$, corresponding to
the ``starts'' of excited subbands, increase with increasing doping level
and agree well with the theoretical calculations, in which the bulk values
of $N_{A}-N_{D}$ are used. This fact also testifies that the disruption of
stoichiometry in surface layers because of probable migration of atoms is
insignificant. A discrepancy with theory is detectable only in the relative
differences (splitting) of occupancies $\Delta
N_{i}/N_{i}=(N_{i}^{-}-N_{i}^{+})/(N_{i}^{-}+N_{i}^{+})$ in the small $N_{s}$
range (see Fig.\,\ref{fig4}). Similar disagreement takes place for inversion
layers on HgCdTe also. The possible reasons for such behavior are discussed
in Ref.\,\cite{Rad96}.


The intensities of Fourier lines for the high-energy branch $I_{i}^{+}$ and
low-energy branch $I_{i}^{-}$ are different (see Fig.\,\ref{fig2} and
Fig.\,\ref{fig7}). At low $N_{i}$, the intensities $I_{i}^{+}$
and $I_{i}^{-}$ are
close for $i=0$ subband, and $I_{i}^{-}>I_{i}^{+}$ for excited subbands. The
ratio $I_{i}^{-}/I_{i}^{+}$ as a function of $N_{s}$ for $i=0$ is shown in
Fig.\,\ref{fig4}. The scattering of the points is due mainly to the fact that
$N_{i}^{\pm }$ values and especially the ratios $I_{i}^{-}/I_{i}^{+}$ depend
strongly (and non-monotonically) on a magnetic field range used in Fourier
transform (for instance, it is clear that Fourier analysis will reveal no
splitting for the range between beat nodes). Nevertheless, a decrease in $%
I_{i}^{-}/I_{i}^{+}$ with increasing $N_{s}$ is clearly visible. It is valid
for excited subbands also.

\subsection{Measurements in tilted magnetic fields}

\begin{figure}
\centerline{\includegraphics[width=10cm]{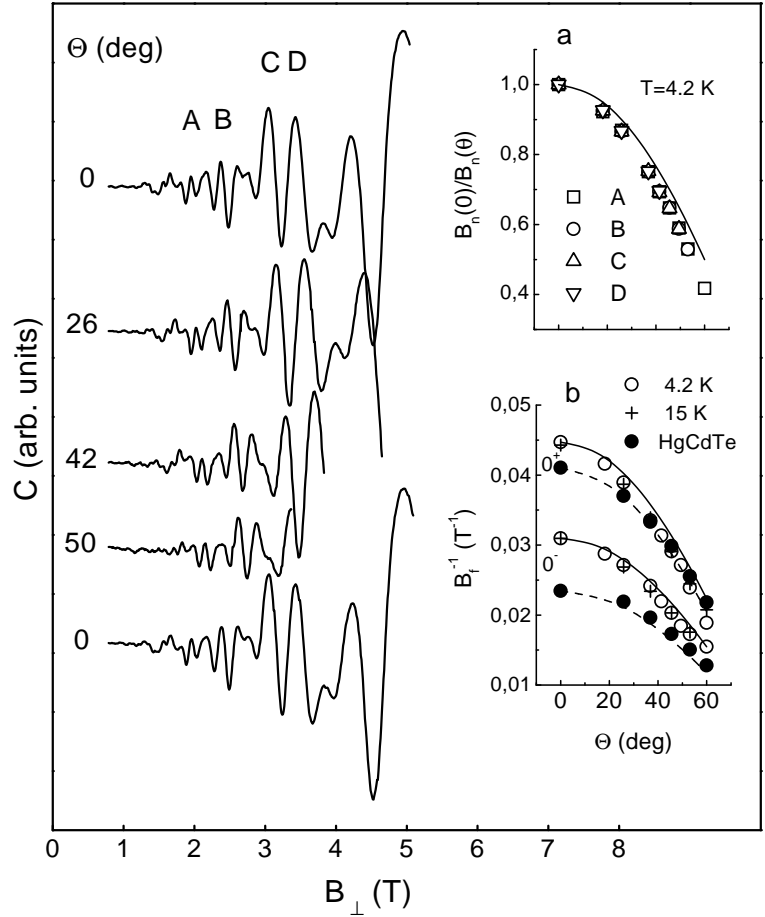}}
\caption{Capacitance oscillations plotted versus the normal component $%
B_{\perp }$ of the applied magnetic field at different angles $\Theta $
between ${\bf B}$ and normal to the 2D layer for sample $S2$ at $%
N_{s}=1.62\cdot 10^{12}$\,cm$^{-2}$. To demonstrate the reproducibility of
the results we plotted two $C(B_{\perp })$ oscillations for $\Theta =0$. The
upper and lower plots are measured before and after angle dependence
measurements respectively. The upper inset shows the angle dependencies $%
B_{n}(0)/B_{n}(\Theta )$ for oscillation maxima marked on the upper $C(B)$
plot. In the lower inset we plotted angle dependencies of fundamental fields
in Fourier transforms for low-energy $(0^{-})$ and high-energy $(0^{+})$
branches of a spectrum for subband $i=0$. The data for inversions layer on
gapless HgCdTe at $T=4.2$\,K are also presented (solid circles and dashed
lines). The lines in the insets are cosine function.}
\label{fig5}
\end{figure}

\begin{figure}
\centerline{\includegraphics[width=10cm]{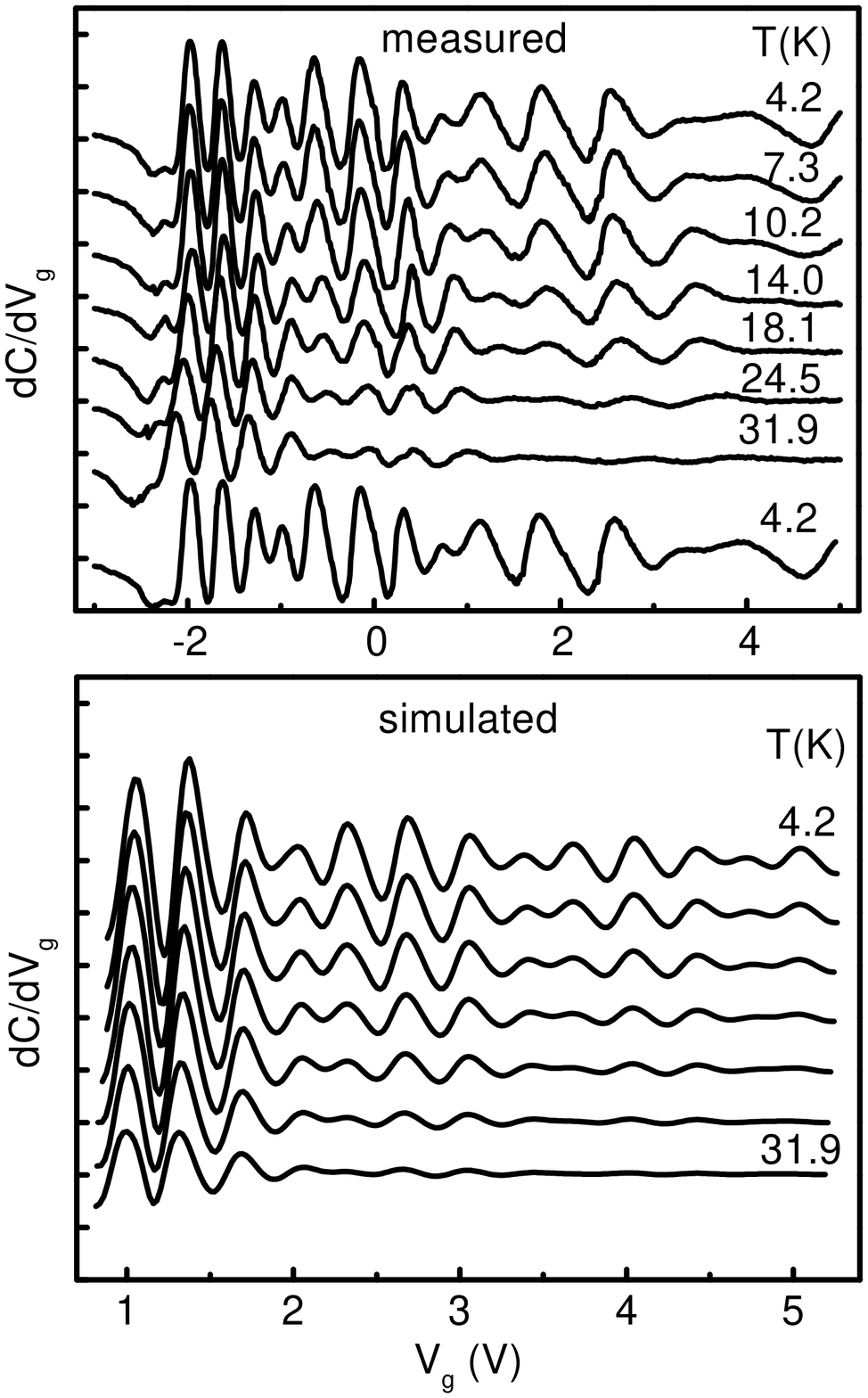}}
\caption{Experimental and simulated temperature evolution of $\frac{dC}{%
dV_{g}}(V_{g})$ oscillations in magnetic field $B=4$\,T for sample $S2$.
Scale division on $\frac{dC}{dV_{g}}$ axis is 50\,pF/V.
The values $T_{N}=10$\,K and $T_{D}=13$\,K are used
in a calculation (we neglected the weak concentration
dependence of the Dingle temperature). The lower curve in the
upper panel is measured two weeks later.}
\label{fig6}
\end{figure}

\begin{figure}
\centerline{\includegraphics[width=12cm]{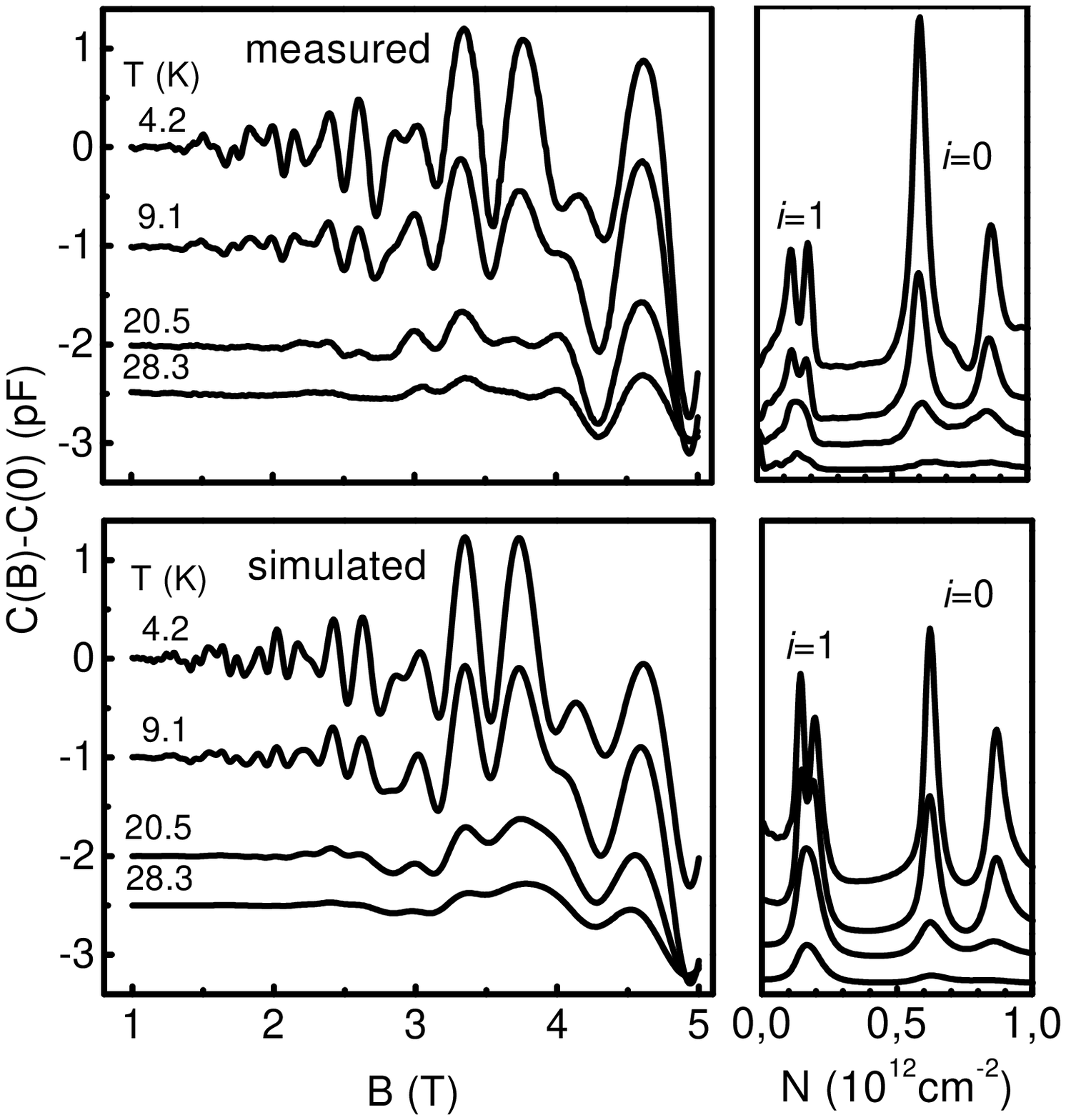}}
\caption{Experimental and simulated temperature evolution of $C(B)$
oscillations and their Fourier spectra for sample $S2$.
The values $T_{D}=12$\,K for $i=0$ and $T_{D}=9.5$\,K
for $i=1$ and $T_{N}=10$\,K are used in the
calculation. The {$C(B)$ }traces are shifted arbitrarily on $C$ axis.}
\label{fig7}
\end{figure}

Although there is no doubt that we are dealing with a 2D system (the
existence of magnetooscillation effect in the capacitance and the
observation of magnetooscillations versus gate voltage in itself testify to
it), the fact experiments in tilted magnetic field were also performed. Some
results for the sample $S2$ are presented in Fig.\,\ref{fig5}.
The magnetic
field positions of the oscillation extrema and the fundament fields in the
Fourier spectra (to a smaller extent) vary only roughly as cosine of the
angle $\Theta $ between ${\bf B}$ and normal to the 2D layer. The clearly
distinguishable deviations from such behavior are observed. Namely, the
experimental angle dependencies are stronger.

There are several reasons for this deviation from classical cosine
dependence, because a number of physical factors are ignored in the
simplified model \cite{Stern}. Firstly, in the strictest sense, such
behavior, even in the case of parabolic dispersion, is valid only for an
ideal 2D system. A condition to be satisfied for cosine dependence is $%
<r>/<z>\gg 1$, where $<r>$ and $<z>$ are the mean sizes of wave function in
the 2D plane and in the confinement direction. In the case of narrow-gap
semiconductors, the width of surface quantum well is relatively large and
such a strong requirement may be not fulfilled (note also that $<z>$ is
energy dependent in this case). In strong magnetic field and at small
surface concentration, the cyclotron radius and the width of 2D layer may be
comparably-sized (especially, for excited subbands) and diamagnetic shift
must weaken the angle dependence. This is contrary to the experimental
behavior. Secondly, the cosine relation is obtained for spinless particles.
This is not the case in a real system. Thirdly, the SO interaction is
neglected in this simple consideration. Undoubtedly, the spinlike effects
can reflect on a spectrum in a tilted magnetic field and modify the angle
dependence. Lastly, the exchange interaction can also make an additional
contribution to the deviation from simple angle dependence.
This assumption has experimental support. For comparison we investigated the
HgCdTe-based samples in a tilted magnetic field. Data for HgCdTe with $%
E_{g}=-95$\,meV and $N_{A}-N_{D}=2\times 10^{17}$\,cm$^{-3}$ are given in
Fig.\,\ref{fig5}. Under the same conditions they
also manifest a deviation from
cosine behavior. However, the deviation is weaker and opposite in sign to
the case of gapless HgMnTe. At the same time, the samples based on HgCdTe
with $E_{g}>0$ show a deviation of the same sign as in HgMnTe, but smaller
in magnitude. Contrary to HgCdTe samples, changes in the structure of
oscillations are observed in HgMnTe inversion layers. Namely, the beat nodes
in oscillations $C(B_{\perp })$ ($B_{\bot }=B\cos \Theta $) are shifted to
the lower LL's numbers with increasing $\Theta $ (i.e., with increasing
total magnetic field $B$) (see Fig.\,\ref{fig5}).

These experimental observations testify that the behavior in tilted magnetic
fields is markedly affected by both SO interaction (which essentially
depends on $E_{g}$ sign, see Sec.\,III. and Fig.\,\ref{fig8}) and exchange
interaction. For narrow-gap semiconductors, the theoretical analysis
requires a consideration of spin from the outset. Strong SO and exchange
interaction and resonant effects lead to serious complication of the
theoretical description even for perpendicular orientation (see Sec.\,III.).
The calculations in tilted magnetic fields are troublesome even for the
simplest parabolic Hamiltonian with a $k$-linear Rashba term. At present we
cannot make A reasonable theoretical analysis of effects in tilted fields we
will restrict our consideration to the case of perpendicular orientation.

\subsection{Temperature effects}

The structure of oscillations and the subband parameters extracted from
oscillations are identical to those in HgCdTe. No features due to exchange
interaction are manifested. Because the exchange effects are determined by a
magnetization and can be varied by the temperature, the investigation of
temperature evolution of oscillations is of prime interest. The results for $%
dC/dV_{g}(V_{g})$ and $C(B)$ oscillations are shown in Fig.\,\ref{fig6} and
Fig.\,\ref{fig7} respectively. As it may be seen, no pronounced changes in
the position of either $dC/dV_{g}(V_{g})$ or $C(B)$ oscillations are
observed. The shift of beat nodes to the high gate voltages and to low
magnetic fields (to the greater LL's numbers) with increasing temperature
(and hence with decreasing magnetization) is the sole temperature effect,
besides the usual diminution of oscillation amplitudes. (Notice, the
direction of shift with the increasing magnetization is similar to the one
observed with increasing total $B$ (at the same $B_{\perp })$ in the tilted
magnetic field experiments). This shift must be attributed to the features
inherent in semimagnetic semiconductors because in HgCdTe-based structures
neither the positions of the oscillations nor those of the beat nodes change
with the temperature.

The analysis based on the Fourier transform of oscillations for different
temperatures cannot yield any information about exchange interaction. On the
other hand, such data cannot be obtained from spin splitting either because,
as noted above, the separate spin components are not observed in the
oscillations at any temperatures. Thus we must settle the question by
capacitance magnetooscillations modeling.

\section{Theoretical Consideration}

\subsection{Landau levels structure}

The direct numerical solution of an initial matrix equation (especially in
the case of non-zero magnetic field) faces obstacles of a fundamental
nature, such as singularity problems and the ambiguity of boundary
conditions on the interface and in the bulk, especially in the case of
gapless semiconductors \cite{Sobk,Godfr}. In this work, we employ a concept
based on the reduction of the matrix equation to a Schrodinger-like equation
with an effective potential \cite{Rad96}. At $B=0$, the results are quite
close to those obtained by numeric calculations
\cite{German,Sobk,Nachev,Ziegler}.
The line of attack of Ref.\,\cite{Rad96}
seems to be a reasonable compromise between accuracy and the ease of the
calculation for a non-zero magnetic field. This simplicity of method is of
considerable advantage for the purposes of oscillations modeling.

Under homogeneous magnetic field ${\bf B}(0,0,B)$ parallel to the direction
of the confinement (surface potential $V=V(z))$ a motion in the 2D plane can
be quantized, using the mean field approximation for exchange interaction.
It can be shown that in the framework of a six-band Kane model with an
allowance for a magnetic field and exchange interaction, the following
matrix equation determines the subband LL's energy $E_{n}(B)$ (we do not
write out the usual expressions, describing the behavior of envelopes in
perpendicular to magnetic field 2D plane) 

%{\small
\begin{equation}
\left[
\setlength{\arraycolsep}{1pt}
\begin{array}{cccccc}
-E_{-}{+}\alpha & \frac{E_{B}\sqrt{3(n-1)}}{2} & \frac{E_{B}\sqrt{n}}{2} & 0 &
0 & s_{b}\hbar \hat{k}_{z} \\
\frac{E_{B}\sqrt{3(n-1)}}{2} & -E_{+}{+}3\beta & 0 & 0 & 0 & 0 \\
\frac{E_{B}\sqrt{n}}{2} & 0 & -E_{+}{-}\beta & s_{b}\hbar \hat{k}_{z} & 0 & 0
\\
0 & 0 & s_{b}\hbar \hat{k}_{z} & -E_{-}{-}\alpha & \frac{E_{B}\sqrt{3(n+1)}}{2}
& -\frac{E_{B}\sqrt{n}}{2} \\
0 & 0 & 0 & \frac{E_{B}\sqrt{3(n+1)}}{2} & -E_{+}{-}3\beta & 0 \\
s_{b}\hbar \hat{k}_{z} & 0 & 0 & -\frac{E_{B}\sqrt{n}}{2} & 0 & -E_{+}{+}\beta
\end{array}
\right] \left(
\setlength{\arraycolsep}{1pt}
\begin{array}{l}
f_{1}^{n-1}(z) \\
f_{3}^{n-2}(z) \\
f_{5}^{n}(z) \\
f_{2}^{n}(z) \\
f_{4}^{n-1}(z) \\
f_{6}^{n-1}(z)
\end{array}
\right) =0,  \label{matrix}
\end{equation}
%}
where $E_{\pm }=E_{n}-V\pm E_{g}/2,$ $s_{b}=\sqrt{\left| E_{g}\right| /2m_{b}%
}$ is Kane velocity, $n$ is LL number. The envelopes $f_{1,4}$ correspond
to $\Gamma _{6}$ symmetry band, $f_{3,6}$ and $f_{2,5}$ to heavy and light
branches of $\Gamma _{8}$ band. ``Magnetic energy'' $E_{B}=\sqrt{%
2m_{b}s_{b}^{2}\hbar \omega _{b}}=\sqrt{2}s_{b}\hbar /\lambda $ ($\hbar
\omega _{b}=\hbar eB/m_{b}c$ is cyclotron energy, $\lambda =\sqrt{c\hbar /eB}
$ is magnetic length) practically does not depend on band parameters because
$s_{b}$ is practically the same for all
Kane semiconductors \cite{Rad87,Johns}.
We denote $\alpha =\frac{1}{2}xN\alpha ^{\prime }\langle
S_{z}\rangle $ and $\beta =\frac{1}{6}xN\beta ^{\prime }\langle S_{z}\rangle
$, where $x$ is the MnTe mole fraction, $N$ is the number of unit cells per
unit volume, $\alpha ^{\prime }$ and $\beta ^{\prime }$ are the exchange
integrals for $\Gamma _{6}$ and $\Gamma _{8}$ bands respectively. The
thermodynamically average $\langle S_{z}\rangle $ of the $z-$ component of a
localized spin $S$ (for Mn$^{2+}$ ions $S=\frac{5}{2}$) can be described via
normalized Brillouin function $B_{S}\left( x\right) $:
\begin{equation}
\langle S_{z}\rangle =-S(1-x)^{18}B_{S}\left( \frac{2\mu _{B}B}{%
k_{B}(T+T_{N})}\right) ,  \label{magnetiz}
\end{equation}
where $T_{N}$ is effective temperature, arising from antiferromagnetic
interaction between Mn$^{2+}$ ions \cite{Bastard,Heiman}.
This factor defines
the magnetic field and temperature dependency of exchange effects.

Resolving the systems (\ref{matrix}) with respect to the components $f_5$
and $f_6$ we obtain the following set of two ``Schrodinger-like'' equations
(for the envelopes $\varphi _5^n=f_5^n/\sqrt{H_n^{+}}$ and $\varphi
_6^{n-1}=f_6^{n-1}/\sqrt{H_n^{-}})$ for the description of the magnetic
spectrum of 2D electrons in surface layers on DMS with inverted bands
(called Kane ``$p$- electrons'' as in Ref.\,\cite{Rad96})

\begin{equation}
\left|
\begin{array}{cc}
\displaystyle \frac{\hbar ^2\hat{k}_z^2}{2m_b}-E_{eff}+U^{+} &
-iU_{so}^{+}-C_g^{+}s_b\hbar \hat{k}_z \\
iU_{so}^{-}+C_g^{-}s_b\hbar \hat{k}_z & \displaystyle \frac{\hbar ^2\hat{k}%
_z^2}{2m_b}-E_{eff}+U^{-}
\end{array}
\right| \left(
\begin{array}{l}
\varphi _5^n \\
\varphi _6^{n-1}
\end{array}
\right) =0  \label{sist2}
\end{equation}

with effective energy
\[
E_{eff}=(E^{2}-m_{b}^{2}s_{b}^{4})/2m_{b}s_{b}^{2},
\]
and effective potential
\[
U^{\pm }=U_{0}+U_{B}^{\pm }+U_{exc}^{\pm }+U_{R}^{\pm }),
\]
in which we single out the following parts: -the spin independent
``Klein-Gordon`` term
\[
U_{0}=(V^{2}-2EV)/2m_{b}s_{b}^{2},
\]
and spin-like terms: ``magnetic potential''
\[
U_{B}^{\pm }=E_{B}^{2}[g^{2}nR^{\pm }+3(n\pm 1)(E_{+}\pm \beta )/(E_{+}\pm
3\beta )]/2m_{b}s_{b}^{2},
\]
''exchange potential''
\[
U_{ex}^{\pm }=[\alpha \beta \pm (\alpha E_{+}+\beta E_{-})]/2m_{b}s_{b}^{2},
\]
and ``resonant'' term describing ``spin-interband`` interaction, arising
from the mixing of $\Gamma _{6}$ and $\Gamma _{8}$ bands by an electric
field
\begin{eqnarray*}
U_{R}^{\pm } &=&\frac{s_{b}^{2}\hbar ^{2}}{2m_{b}s_{b}^{2}}\left[ \frac{3}{4}%
\left( \frac{1+L_{n}^{\pm }}{H_{n}^{\pm }}\right) ^{2}+\frac{L_{n}^{\pm }}{%
H_{n}^{\pm }(E_{+}\pm 3\beta )}\right] \left( \frac{dV}{dz}\right) ^{2} \\
&&+\frac{s_{b}^{2}\hbar ^{2}}{4m_{b}s_{b}^{2}}\frac{(1+L_{n}^{\pm })}{%
H_{n}^{\pm }}\frac{d^{2}V}{dz^{2}},
\end{eqnarray*}
where the following designations are used:
\[
L_{n}^{\pm }=3E_{B}^{2}(n\pm 1)/4(E_{+}\pm 3\beta )^{2},\ H_{n}^{\pm
}=E_{-}\pm \alpha -L_{n}^{\pm }(E_{+}\pm 3\beta ),
\]
\[
R_{n}^{\pm }=H_{n}^{\pm }/H_{n}^{\mp },\ g=-1
\]
The spin-orbit terms and the coefficients at linear in $\hat{k}_{z}$ terms
are determined by
\[
U_{so}^{\pm }=C_{g}s_{b}\hbar \sqrt{R_{n}^{\mp }}\left[ \frac{1+L_{n}^{\pm }%
}{H_{n}^{\pm }}+\frac{1+L_{n}^{\mp }}{2H_{n}^{\mp }}(R_{n}^{\pm }-1)\right]
\frac{dV}{dz},
\]
\[
C_{g}^{\pm }=C_{g}\sqrt{R_{n}^{\mp }}(R_{n}^{\pm }-1),\ C_{g}=g\frac{%
E_{B}\sqrt{n}}{4m_{b}s_{b}^{2}}.
\]
The second term in an expression for $U_{B}^{\pm }$ and the dimensionless
parameter $L_{n}^{\pm }$ arise from the interaction with heavy hole branch.
It must be stressed that the exchange interaction causes not only the
appearance of an exchange term in the effective potential, but also a
modification of the terms, describing the ''resonant'' and\ SO interaction.

The LL's of 2D electrons in Kane semiconductors with $E_{g}>0$ (``$s$%
-electrons'') are described by the same set (\ref{sist2}) (for the envelopes
$\varphi _{2}^{n}=f_{2}^{n}/\sqrt{H_{n}^{+}}$ and $\varphi
_{1}^{n-1}=f_{1}^{n-1}/\sqrt{H_{n}^{-}}$) but with $g=+1$, $L_{n}^{\pm }=0$,
$H_{n}^{\pm }=E_{+}\pm \beta $. It can be shown that the equations for Dirac
like electrons in a magnetic field are the same as for Kane ``s- electrons''
but with $g=+2$. It should be emphasized that unlike the $B=0$ case
\cite{Rad96}, the set (\ref{sist2}) cannot be separated and reduced to
independent equations for individual spin components because of the SO
interaction $U_{so}^{\pm }$ (for ``$p$- electrons'', because of linear in $%
\hat{k}_{z}$ terms also).

From this point on, we shall restrict our consideration to the semiclassical
approximation just as in quantization of spectrum described by Eqs. (\ref
{sist2}) (it is clear that the use of the semiclassical approximation
immediately in (\ref{matrix}) results in the loss of spinorlike effects), so
also in calculation of the surface potential $V(z)$. A semiclassically
self-consistent potential in such treatment is calculated in the frame of
quasirelativistic modification of the Thomas-Fermi method. The validity of
such an approach in narrow-gap semiconductors was argued and demonstrated by
a comparison with numerical self-consistent calculations in many papers
\cite{Ando85,Rad88,Paasch} (see also Ref.\,\cite{Rad96} and references
therein). Substituting as usual $\varphi _{i}^{m}=C_{i}^{m}\exp (i\int
k_{z}(z)dz)$ and neglecting the proportional to $i(zdk_{z}/dz+k_{z})^{2}$
terms (higher-order terms in the expansion of the action in powers of $\hbar
)$ we obtain from (\ref{sist2}) the quasiclassical expression for
``spin-split'' $z-$components of wave vector
\begin{equation}
k_{z}^{\pm } =\frac{\sqrt{2m_{b}s_{b}^{2}}}{s_{b}\hbar }\{K\mp \lbrack K^{2}-
(E_{eff}-U^{+})(E_{eff}-U^{-})+
U_{so}^{+}U_{so}^{-}]^{\frac{1}{2}}\}^{\frac{1}{2}}  \label{Kz}
\end{equation}
with
\[
K=E_{eff}-(U^{+}+U^{-})/2-m_{b}s_{b}^{2}C_{g}^{2}(R_{n}^{+}-1)(R_{n}^{-}-1).
\]
Together with the Bohr-Sommerfeld quantization rule
\begin{equation}
\int\nolimits_{V(z=0)}^{V(k_{z}=0)}k_{z}(E,V)\left( \frac{dV}{dz}\right)
^{-1}dV=\pi (i+\frac{3}{4})  \label{Bor}
\end{equation}
they define the magnetic levels $E_{n}^{\pm }(i,B)$ in surface quantum well $%
V(z)$.

For the Dirac-like electrons, the derived equations may be considered as a
generalization of the result of Ref.\,\cite{Ritus} that allows for
magnetic quantization and spin-like effects. In the case of Kane ``$s-$
electrons`` and without exchange interaction, the Eqs.\,(\ref{sist2}) and (%
\ref{Kz}) coincide with the corresponding expressions
in Ref.\,\cite{Uem}. As might be expected, at $\alpha =0,\ \beta=0,\
n\rightarrow \infty,\ E_{B}\sqrt{n\pm 1}$ and $E_{B}\sqrt{n}\rightarrow
s_{b}\hbar k_{s}$ ($k_{s}$ is 2D wave vector) the Eq. (\ref{Bor}) with $%
k_{z} $ from Eq. (\ref{Kz}) is reduced to the corresponding equation
in Ref.\, \cite{Rad96}, describing subband dispersions $E_{i}^{\pm }(k_{s})$ at $%
B=0$. Lastly, the Eq. (\ref{Kz}) at $V(z)=const$ determines the Landau
subbands $E_{n}^{\pm }(B,k_{z})$ in the bulk of Kane (at $\alpha =0$ and $%
\beta =0$ non-magnetic) semiconductors.

\begin{figure}
\centerline{\includegraphics[width=10cm]{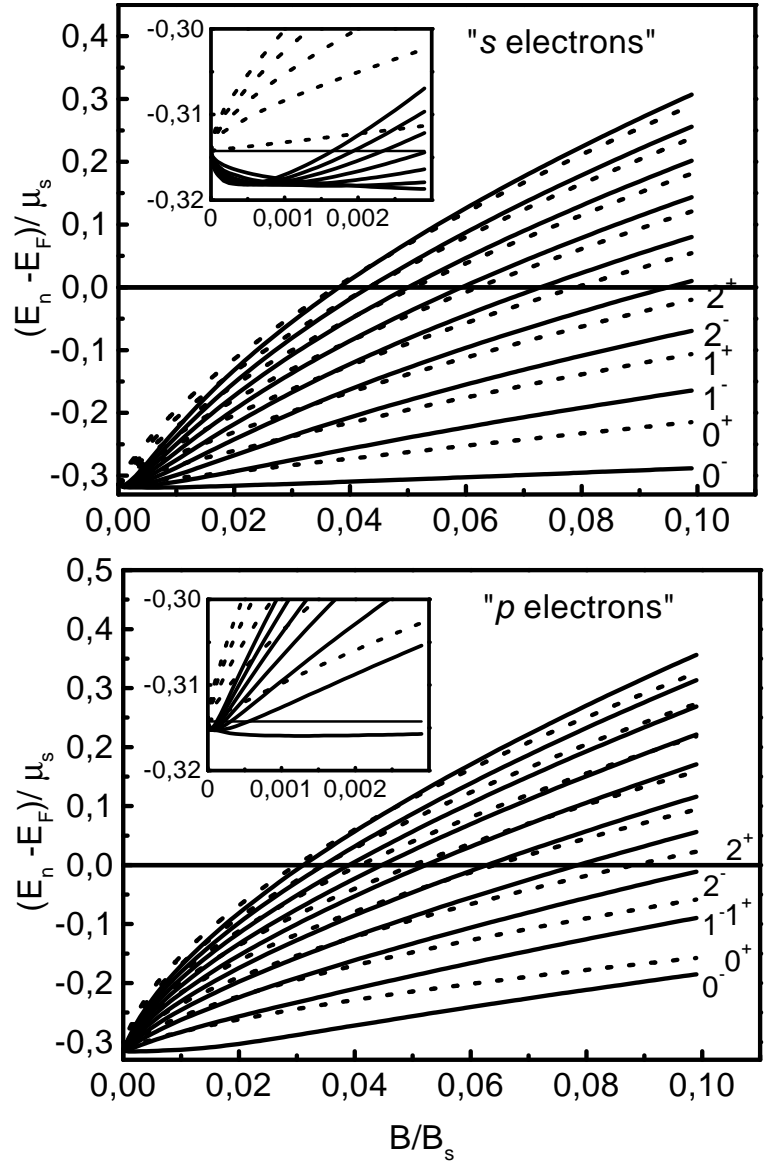}}
\caption{A spectrum in a magnetic field for Kane ``$s$ '' and ``$p$
electrons'' in dimensionless units in the pseudoultrarelativistic limit $%
E_{g}=0$. Drastic spin-orbit perturbation of a magnetic spectrum as compared
with an ordinary set of Landau levels is observed at the energies near the
subband bottom [shown in insets] as well as nearby the Fermi level. The
effect is essentially different for materials with direct (upper panel) and
inverted (lower panel) bands. }
\label{fig8}
\end{figure}

It must be noted that the proportional to $C_{g}^{2}$ term in expression for
$K$ (arising from linear in $\hat{k}_{z}$ terms in (\ref{sist2})) has little
effect in comparison with the spin-orbit term in Eq. (\ref{Kz}). However,
ignoring this term in calculations at $U_{so}^{\pm }=0$ can introduce large
error up to the inversion of the order of spin sublevels. In
pseudo-ultrarelativistic limit $E_{g}=0$ and without exchange interaction,
the spectra in magnetic fields for both ``$s$ '' and ``$p$-electrons'' are
scale invariant with respect to the surface band bending $\mu _{s}$ as well
as in the case of $B=0$ \cite{Rad96,Rad87}. The results for ground subband
are plotted in Fig.\,\ref{fig8} in dimensionless (normalized to $\mu _{s}$)
coordinates $E/\mu _{s}-B/B_{s}=(E_{B}/\mu _{s})^{2}$ $(B_{s}=c\mu
_{s}^{2}/2es_{b}^{2}\hbar )$. It is seen that the SO interaction leads to so
drastic a reconstruction of 2D spectrum in magnetic fields, that the
description of spin splitting by such a non-relativistic parameter as $g$%
-factor loses in essence its physical meaning. The SO splitting far exceeds
a contribution due to exchange interaction (in the scale of Fig.\,\ref{fig8},
the changes induced by exchange interaction are practically
indistinguishable for reasonable exchange coupling parameters). This is true
for narrow-gap DMS with $E_{g}>0$ also. In connection with this, the results
of the analysis of 2D systems in asymmetrical quantum wells in these
materials are to be revised, because they ignore the SO interaction.

\subsection{Density of states and capacitance in magnetic field}

For 2D systems with the multisubband spectrum, the differential capacitance
of space charge region with 2D electron gas is determined by the sum $%
C_{sc}=\sum C_{i\sigma }$ of ``partial'' subband capacitances
\begin{equation}
C_{i\sigma }=e^{2}\frac{dN_{i\sigma }}{d\mu _{s}}=\frac{e^{2}}{\pi
s^{2}\hbar ^{2}}\frac{d\mu _{Fi}}{d\mu _{s}}\frac{d}{d\mu _{Fi}}\int
D_{i\sigma }(E)f(E-\mu _{Fi})dE,  \label{cap_i}
\end{equation}
where $N_{i\sigma }$ is the surface concentration in spin branch $\sigma
=\pm $ of $i$-th subband, $f(E-\mu _{Fi})$ is the Fermi-Dirac distribution
function and $\mu _{Fi}$ is the subband Fermi energy. The
magnetooscillations of capacitance are dictated by peculiarities in the DOS $%
D_{i\sigma }(E)$ in a magnetic field, i.e. by the energy position and
broadening of LL's. The rigorous quantitative analysis of disorder broadened
LL's is a complicated problem even in one-band approximation
\cite{Ando82,Spies}. We shall perform the treatment by neglecting the mixing
between LL's and assuming a Gaussian shape of each level \cite{Ando74,Gerh}
\begin{equation}
D_{i\sigma }=D_{i}^{\pm }(E)=\frac{eB}{c\hbar \sqrt{2\pi ^{3}}}%
\sum\limits_{n=0}^{\infty }\frac{1}{\Gamma _{i}^{\pm }}\exp \left[ -2\left(
\frac{E-E_{ni}^{\pm }}{\Gamma _{i}^{\pm }}\right) ^{2}\right]  \label{dens1}
\end{equation}
with broadening parameters $\Gamma _{i}^{\pm }$, independent of LL's number,
but dependent on $B$ and (for discussed non-parabolic system) on energy. The
analytical description of magnetooscillation phenomena, if possible, is
preferable because it is more transparent, and easy to interpret in contrast
to direct numerical methods. In addition, the peculiar features of
oscillation phenomena in 2D systems with quasirelativistic spectra as
compared with those of parabolic spectra \cite{Ando74} can thus be
clarified. For this purpose we approximate the spin-split subband
dispersions by analytical relativistic-like expression with the subsubband
masses $m_{0i}^{\pm }$ and velocities $s_{i}^{\pm }$. Such approximation
holds for practically all the energy range of interest. The replacement in
Eq.(\ref{dens1}) of numerical solutions of Eq. (\ref{Bor}) with $k_{z}$ from
Eq. (\ref{Kz}) by the corresponding quasiclassical spectrum in a magnetic
field (for simplicity we shall drop the spin subband indices $\sigma =\pm $
in the following)
\begin{equation}
E_{ni}=\sqrt{E_{Bi}^{2}\left( n_{i}+\delta _{i}\right) +m_{0i}^{2}s_{i}^{4}}%
-m_{0i}s_{i}^{2}  \label{disp_B}
\end{equation}
($E_{Bi}=\sqrt{2}s_{i}\hbar /\lambda $) does not noticeably reflect on the
shape of the DOS. As to the LL's energy positions (and the positions of
oscillations on $B$), their values are to be calculated on the basis of Eqs.
(\ref{Kz}) and (\ref{Bor}). In the Born approximation for short-range
scattering, the parameter $\Gamma _{i}$ is related to the classical
scattering time in zero magnetic field $\tau _{i}$
by $\Gamma _{i}^{2}=\sqrt{%
2/\pi }\hbar ^{2}\omega _{ci}(E)/\tau _{i}$ ($\omega _{ci}(E)=eB/cm_{ci}(E)$,
$m_{ci}(E)=m_{0i}+E/s_{i}^{2}$) \cite{Ando82}. Inserting the above
expressions for $E_{ni}$ and $\Gamma _{i}$ in (\ref{dens1}) and using a
Poisson summation formula, we arrive (under the assumptions $\Gamma
_{i}<<E_{Bi}$ and $\Gamma _{i}<<E+m_{0i}s_{i}^{2}$, which are justified
even for the states nearby the subband bottom) at a ``harmonic''
representation for DOS (convenient for the description of oscillation
effects)

\begin{equation}
D_{i}(E,B)=D_{i}(E,0)[1+2\sum\limits_{j=1}^{\infty }(-1)^{j}\exp \left( -%
\frac{j^{2}\pi }{\omega _{ci}(E)\tau _{i}}\right) \cos (2\pi jn_{i}(E))],
\label{dens3}
\end{equation}
where $D_{i}(E,0)=m_{di}(E)/2\pi \hbar ^{2}=(E+m_{0i}s_{i}^{2})/2\pi
s_{i}^{2}\hbar ^{2}$ is DOS at $B=0$. Notice, that in the case of a
two-dimension system, as is easy to show, the effective mass of DOS $m_{di}$
coincides with the cyclotron effective mass $m_{ci}$ for any dispersion law.
The LL ``number'' $n_{i}(E)$ in Eq. (\ref{dens3}) is regarded as an
arbitrary quantity (not necessarily an integer) and is determined at a given
energy by the solutions of Eqs. (\ref{Kz}) and (\ref{Bor}). One can readily
see that in ``non-relativistic'' limit ($s_{i}\rightarrow \infty )$, Eq. (%
\ref{dens3}) is reduced to the Ando formula \cite{Ando74} if the terms with $%
j>1$ are neglected. It has been shown that for all parameters and regimes of
practical interest, the DOS given by (\ref{dens1}) and (\ref{dens3}) are
practically equivalent. This is true in the case of both the sinusoidal
shape of DOS (when only ground harmonic predominates in sum (\ref{dens3}))
and the nonsinusoidal shape of $D(E)$ dependence. The latter case takes
place at weakly broadened LL's, when the individual spin components can be
resolved in total DOS.

Substituting Eq. (\ref{dens3}) in Eq. (\ref{cap_i}) and neglecting the
integrals of odd functions (this is justified at $kT<<\mu _{Fi}$ ) we obtain
an expression for magnetocapacitance
\begin{eqnarray}
\frac{C_{i}(B)}{C_{i}(0)} &=& 1-2\sum\limits_{j}\left( -1\right) ^{j}\exp
\left( -\frac{j^{2}\pi }{\omega _{ci}\tau _{i}}\right)
\int\limits_{0}^{\infty }\frac{\cos (jb_{i}y)}{2\cosh ^{2}(y/2)}  \nonumber
\\
&\times& \{\cos (2\pi jn_{i})[\cos (2\pi jc_{Bi}^{2}y^{2})-(jc_{\tau
i}+yc_{Ti}\tan (jb_{i}y))\sin (2\pi jc_{Bi}^{2}y^{2})]  \\
&-&\sin (2\pi jn_{i})[\sin (2\pi jc_{Bi}^{2}y^{2})+(jc_{\tau i}+yc_{Ti}\tan
(jb_{i}y))\cos (2\pi jc_{Bi}^{2}y^{2})]\}dy,  \nonumber \label{oscgen}
\end{eqnarray}
where
\[
y=(E-\mu _{Fi})/kT,\omega _{ci}=\omega _{ci}(\mu _{Fi}),b_{i}=2\pi kT/\hbar
\omega _{ci},
\]
\[
c_{Ti}=K_{mi}\frac{kT}{\mu _{Fi}},\ c_{\tau i}=K_{mi}\frac{\hbar }{%
2\tau _{i}\mu _{Fi}},\ c_{Bi}=\frac{kT}{E_{Bi}}=\frac{kT\lambda }{%
\sqrt{2}s_{i}\hbar },
\]
\[
K_{mi}=\left[ 1+\frac{d(m_{0i}s_{i}^{2})}{d\mu _{Fi}}\right] \left[ 1+\frac{%
m_{0i}s_{i}^{2}}{\mu _{Fi}}+\frac{d(m_{0i}s_{i}^{2})}{d\mu _{Fi}}\right]
^{-1},
\]
and

\[
C_{i}(0)=e^{2}\left[ D_{i}(\mu _{Fi},0)+\frac{\mu _{Fi}}{2\pi s_{i}^{2}\hbar
^{2}}\frac{d(m_{0i}s_{i}^{2})}{d\mu _{Fi}}\right] \frac{d\mu _{Fi}}{d\mu _{s}%
}.
\]

Note that the identity between the capacitance in a zero magnetic field $%
C_{i}(0)$ and the quantity $e^{2}D(\mu _{F},0)d\mu _{F}/d\mu _{s}$ breaks
down in 2D systems with relativistic-like spectra \cite{Rad89a}. As compared
to the semiconductors with standard bands an expression (\ref{oscgen})
contains new parameters of the theory $c_{T}$, $c_{\tau }$ and $c_{B}$. The
first and the second are due to the energy dependence of $m_{di}$ in an
expression for DOS at $B=0$ and of $m_{ci}$ in exponential (Dingle) factor
in Eq. (\ref{dens3}). This energy dependence is the result of two effects.
The first is the non-parabolicity of subband dispersions. The second is due
to the change of the parameters of subsubband dispersions (basically $%
m_{0i}^{\pm }(\mu _{s})$ ) under the modulation of surface potential. The
parameter $c_{B}$ is determined in fact only by the ratio $T/\sqrt{eB}$ and
is independent of subband parameters, because $s_{i}^{\pm }$ differs only
slightly from the ''universal'' value of $s_{b}$. (The arising of a $c_{B}$
parameter for weak relativistic Fermi gas $\mu _{F}<ms^{2}$ of Dirac
electrons is reported in Ref.\,\cite{Vshiv}).

For strong enough magnetic field, the faze $2\pi jc_{B}^{2}y^{2}$ is close
to zero in the range of small $y$ (this range gives a dominant contribution
to the integral in Eq. (\ref{oscgen})) and Eq. (\ref{oscgen}) can be
evaluated as
\begin{equation}
\frac{C_{i}(B)}{C_{i}(0)} \approx 1-2\sum\limits_{j}\left( -1\right) ^{j}%
\frac{j\pi b_{i}}{\sinh (j\pi b_{i})}\exp \left( -\frac{j^{2}\pi }{\omega
_{ci}\tau _{i}}\right) 
\times \left[ \cos (2\pi jn_{i})-jc_{\tau i}\sin (2\pi jn_{i})\right]
\label{oscapr}
\end{equation}
We point out that in the two-dimension case, the argument of the Dingle
exponent is quadratic in harmonic number $j$. In ''non-relativistic'' limit $%
s_{i}\rightarrow \infty $, the values $c_{Ti}=c_{\tau i}=c_{Bi}=0$ and Eq. (%
\ref{oscapr}) reduces to the Ando expression for oscillations in 2D systems
with parabolic spectra \cite{Ando82,Ando74}. Thus the parabolic approach is
appropriate in the case of strong enough magnetic fields $E_{Bi}>(5\div
10)kT $ and not-too-broadened LL's $(\hbar /2\tau _{i}\leq \mu _{Fi})$.

\section{Results of modeling and discussion}

The capacitance of the MOS structures $%
C(B)=C_{ox}C_{sc}(B)/(C_{ox}+C_{sc}(B))$ was calculated using the value of
oxide capacitance $C_{ox}$, determined from the capacitance in a strong
accumulation regime. The change of charge in the depletion layer with $\mu
_{s}$ in the inversion band bending range (in narrow-gap semiconductors such
changes can be quite significant) is taken into account in the calculation.
As a rule, the theoretical capacitances $C(0)$ are in good agreement with
the ones measured at the same surface density. In a modeling, the surface
potential and subband Fermi energies are supposed to be constant when a
magnetic field is changing. The alternative model is based on the assumption
that the surface density is fixed. However, both models give
indistinguishable results at large enough LL's broadening (this is
manifested by the cosine form of experimental oscillations) \cite{Bass}. The
temperature dependencies of band parameters and bulk Fermi energy are
accounted for in the calculations.

Although the general shapes of simulated and measured oscillations $C(B)$
are well matched, the exact magnetic field positions of the peaks and beat
nodes are somewhat different. This is not surprising, because a number of
physical factors are ignored or cannot be exactly taken into account in a
theory (the contribution of remote bands, interface contribution to the SO
interaction (see below), the deviation of real surface potential and Landau
level shape from those calculated, the superposition of oscillations from
different subbands). At the same time the positions of oscillations and
especially beat nodes are very sensitive to each of these factors. The
adjustable phase factor (a correction $\Delta n_{i}$ to LL's ``number'' $%
n_{i}$ in the expressions (\ref{dens3}), (\ref{oscgen}) and (\ref{oscapr}))
was introduced for reasons of convenience for a comparison of the
temperature evolution in the measured and calculated oscillations. Its
magnitude was chosen to fit the high-field node position of the beat pattern
at T=4.2\,K. Any physically meaningful results discussed are not affected by
the choice of this factor.

The oscillations calculated with this correction and their Fourier
transforms are plotted in Fig.\,\ref{fig7}. The agreement is quite good with
respect to the structure of oscillations as well as the amplitudes. However,
a distinguishable difference in the ``number'' of oscillations between beat
nodes for measured and calculated plots is observed. These results, as well
as the similar data on $dC/dV_{g}(V_{g})$ oscillations (see Fig.\,\ref{fig6}),
testify to the small (but distinguishable) underestimation of SO splitting
by the theory. This conclusion is valid for HgCdTe also. Note that a
treatment based on the analysis of Fourier spectra does not give a clearly
detectable discrepancy between experiment and theory (excluding the small
$N_{s}$ range (see Fig.\,\ref{fig4})). This inconsistency with theory can be
caused by the interface contribution to the SO interaction \cite{Engels},
which cannot be treated in the framework of effective mass method. In
accordance with the experiment, the individual spin components are not
exhibited in simulated $C(B)$ or $dC/dV_{g}(V_{g})$ oscillations even for
the lowest LL's at any reasonable broadening parameters and magnetic fields
of experimental interest. The results of modeling based on Eq. (\ref{oscgen}%
) do not differ from the results obtained in the approximation (\ref{oscapr}%
) practically for all $N_{s}$, $T$ and $B$ at which the oscillations are
observed experimentally. As a rule, the contribution from the sine term in
Eq. (\ref{oscapr}) is small.

\subsection{Exchange interaction effects}

A modeling shows that the magnetic field positions of the oscillations
beyond the neighborhood of beat nodes are unaffected by the exchange effects
even at the lowest temperatures. As a result these positions are temperature
independent, as occurs experimentally. This is true for any available values
of the exchange constants $N\beta ^{\prime }$ and $N\alpha ^{\prime }$
(literature data vary markedly,
see Refs.\,\cite{Furd,Mink,Mink2} and \cite{Furd88}
and references in these works; note that in different
papers the notations for exchange constants differ in sign \cite{Furd}). As
noted in Sec.\,II, an exchange interaction is very weakly manifested in the
studied system, showing itself as only a slight temperature shift of beat
nodes. Because the oscillation amplitudes in the neighborhood of nodes are
small even at $T=4.2$\,K and they decrease drastically with temperature, the
narrow range of $T<10\div 15$\,K is accessible to the quantitative analysis.
Thus the results are not critically sensitive to a choice of $N\beta
^{\prime }$ and $N\alpha ^{\prime }$. Secondly, the rate of shift depends on
the product of the exchange parameters $N\beta ^{\prime }$ and $N\alpha
^{\prime }$ and the magnetization $\langle S_{z}\rangle $. So the variations
in $N\beta ^{\prime }$ and $N\alpha ^{\prime }$can be cancelled out by the
variation in $T_{N}$, which is used as an adjustable parameter (see below).
It must be stressed that the terms containing a parameter $\beta $ play the
dominant role in Eqs. (\ref{matrix}), (\ref{sist2}) and (\ref{Kz}) for $p$%
-electrons. Under the conditions of experimental interest, the temperature
shift of beat nodes is also slightly sensitive to the variations $N\alpha
^{\prime }$ in a wide range even if a small $N\beta ^{\prime }$ is chosen.

Although we performed the calculations for a different net of exchange
parameters, the results discussed in this Section correspond to $N\beta
^{\prime }=1.5$\,eV and $N\alpha ^{\prime }=-0.4$\,eV, unless otherwise
specified. The results are only slightly sensitive to variances of $N\beta
^{\prime }$ in the range $1.35\div 1.65$\,eV and do not differ at all for $%
N\alpha ^{\prime }$ varied through $-(0.35\div 0.50)$\,eV range. These values
are close to those obtained in Refs.\cite{Mink} and \cite{Mink2}
for narrow-gap and gapless HgMnTe with small $\left| E_{g}\right| $ by the
tunnel spectroscopy method. We suppose that these data (similar values for
gapless HgMnTe have been obtained in many works (see Refs.\,\cite{Furd,Mink}
and \cite{Mink2} and references in these
works) are more suitable for the purposes of this work, because in studied
surface quantum wells the typical electron energies are of the order of or
even considerably more than $\left| E_{g}\right| $. In tunnel experiments,
the LL's energy positions of ``$p$- electrons'' as a function of magnetic
field are measured at energies up to 150\,meV. At the same time, the states
with the energy near band bottom are tested by the traditional methods.

It should be noted in connection with this that a decrease of $|N\alpha
^{\prime }|$ for $s$-electrons in in wide-gap CdMnTe-CdMgMnTe quantum well
with increasing energy is reported in a recent paper Ref.\,\cite{Merkul}%
. The effect is attributed to the admixing of $\Gamma 8$ band states to $%
\Gamma 6$ band at finite ${\bf k}$-vectors which leads to switching-on of a
kinetic exchange for electrons of $\Gamma 6$ band with the $d$ electrons of
Mn ions. Note that in narrow-gap semiconductors, the interband mixing,
described by the Kane Hamiltonian (\ref{matrix}), results in a strong (and
energy dependent) contribution of the $N\beta ^{\prime }$ containing terms
to a spectrum of $\Gamma 6$ band. This is true without allowing for the
energy dependence of parameter $N\alpha ^{\prime }$. As for electrons of $%
\Gamma 8$ band (``$p$- electrons``), the value of the exchange parameter ($%
N\beta ^{\prime }$) is from the outset governed mainly by kinetic exchange
(at any ${\bf k}$-vector). In this case, an increase of ${\bf k}$-vector
cannot play a critical role. At present, we do not have evidence of energy
dependence of parameter $N\beta ^{\prime }$. The absence of an essential
change in the value of $N\beta ^{\prime }$ is noted
in Refs.\,\cite{Merkul} and \cite{Bhatt}. Together with the weak
sensitivity of the observed effects to a choice of $N\beta ^{\prime }$ and $%
N\alpha ^{\prime }$, this suggests that any possible energy dependence of
exchange parameters cannot markedly reflect on the results.

Once the exchange parameters are chosen, two parameters can be obtained when
the modeling fits experimental data: the effective temperature $T_{N}$,
which describes the temperature shift of beat nodes, and the Dingle
temperature $T_{D}=\hbar /k_{B}2\pi \tau $ (used by us as the characteristic
of the scattering instead of collision time $\tau $), which determines the
oscillation amplitudes.

\begin{figure}
\centerline{\includegraphics[width=10cm]{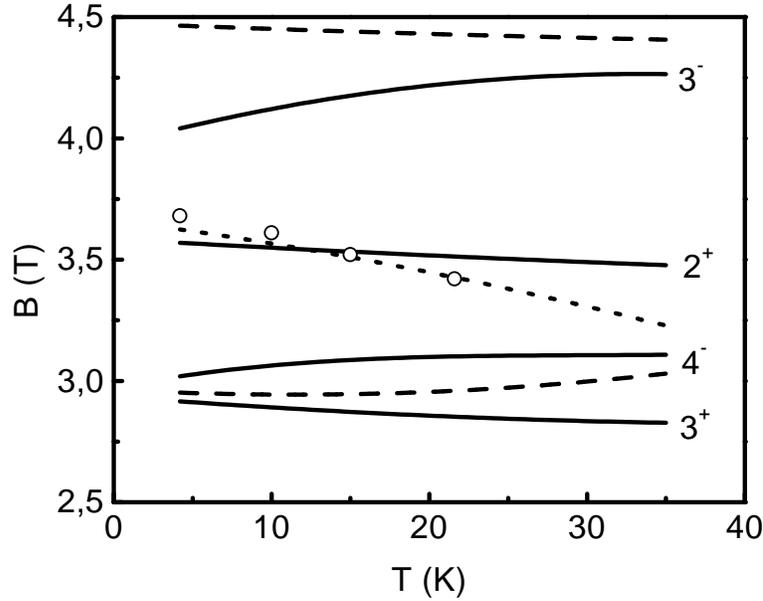}}
\caption{Magnetic field positions of the Landau levels (full curves), the
maxima of simulated oscillation (broken curves) and the beat node (dotted
curve) calculated as a function of the temperature for ground subband at
subband occupancy $N_{0}=0.59\times 10^{12}$\,cm$^{-2}$ (sample $S2$). $%
T_{N}=10$\,K. The experimental positions of the beat node are shown as
points.}
\label{fig9}
\end{figure}

\begin{figure}
\centerline{\includegraphics[width=10cm]{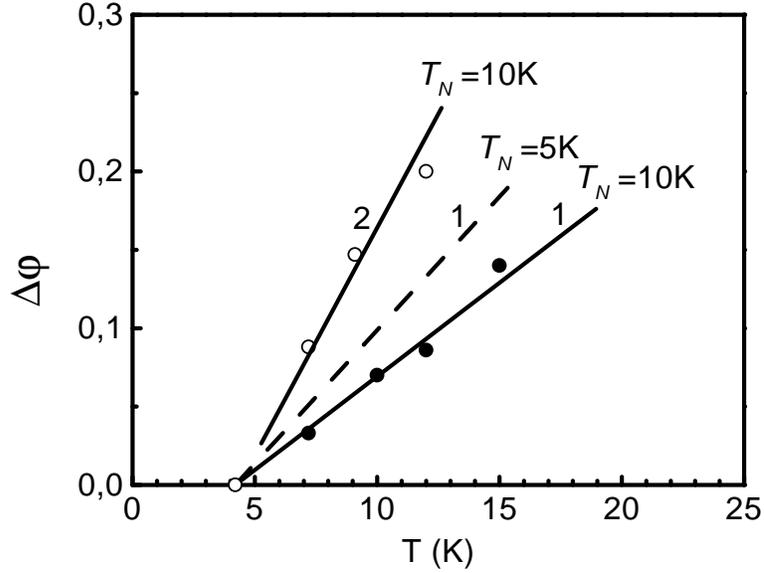}}
\caption{Calculated (curves) and measured (points) temperature shift of
magnetic field position of the beat node (in the units of fundamental
period) at $N_{0}=1.45\times 10^{12}$\,cm$^{-2}$ (curves 1) and
$N_{0}=0.59\times 10^{12}$\,cm$^{-2}$ (curve 2) for sample $S2$.}
\label{fig10}
\end{figure}

Although a decrease of magnetization with increasing temperature results in
the slight energy shift of spin sublevels, the position of resulting
oscillations on a magnetic field is almost unchanged (see Fig.\,\ref{fig9}).
The rate of shift with temperature depends on the $N_{i}$ and node number,
as is shown in Fig.\,\ref{fig10}. At the same time, the value of $T_{N}$,
extracted from the fit of the temperature evolution of oscillations, is
almost the same for different nodes and $N_{i}$, which counts in favor of
model used. The results of the simulation are not critically dependent on
the exact value of $T_{N}$ chosen. However, the ``best fitting'' value
$T_{N}=10\pm 1.5$\,K must be a fairly good estimation. Unfortunately,
as far as we know, the low-temperature data on $T_{N}$ value for bulk HgMnTe
with $x=0.04$ are absent. Most of the literature data are obtained either
for high temperatures or for samples with Mn content $x\leq 0.025$. However
the value $T_{N}=10$\,K does not contradict other published data. If the
sample-independence of spin-spin interaction is postulated, $T_{N}$ is
nearly proportional to $x(1-x)^{18}$ (see Ref.\,\cite{Bastard}). Using
the low temperature data from Ref.\,\cite{Bastard} for a sample with $%
x=0.01$ ($T_{N}=2.9$\,K at $T=2$\,K ) we can estimate the value of $T_{N}$ for
samples with $x=0.04$ as $T_{N}\approx 8$\,K. This is somewhat less than the
measured value, but $T_{N}$ can also be temperature dependent. \cite{Dobr}
For example, for the same sample with $x=0.01$, $T_{N}$ is equal to 7\,K in
a high temperature range \cite{Bastard}. It must be noted that the above
estimations are based on assumptions which can be violated (including a
phenomenological expression itself (\ref{magnetiz})) for $x>0.02$ and low
temperature.

The SO interaction not only contributes predominantly to the spin splitting
but also suppresses the splitting due to exchange interaction. As an
example, the SO splitting, corresponding to the first beat node
in Fig.\,\ref{fig7}, is 24.9\,meV. If we take exchange interaction into account, the
splitting increases by only 3.7\,meV even at T=4.2\,K.
At the same time, the
exchange splitting, calculated without allowance for SO interaction,
is 6.4\,meV. That is why the exchange effects show themselves
only as a weak change
in the structure of oscillations near the beat nodes, where the oscillations
from different spin branches quench each other.

Let us now return to the dependence of observed exchange effects (the
temperature shift of beat nodes) on the value of exchange parameters. Only
the shift of beat nodes to low LL's numbers is observed at low temperatures
with decreasing $N\beta ^{\prime }$ (the shift is slightly sensitive to the
variations $N\alpha ^{\prime }$ in $-(0.25\div 0.5)$\,eV range). As a result,
the rate of temperature shift of nodes decreases and becomes less than the
one calculated at $N\beta ^{\prime }=1.5$\,eV. However, at $N\beta ^{\prime
}>0.75$\,eV, this decrease can be cancelled out by a decrease in $T_{N}$. For
$N\beta ^{\prime }=1.0$\,eV and $N\alpha ^{\prime }=-0.4$\,eV the shifts
coincide with those found for $N\beta ^{\prime }=1.5$\,eV and $N\alpha
^{\prime }=-0.4$\,eV if the value $T_{N}=4$\,K is chosen. The oscillations in
both cases are practically the same at all $B$ (including the ranges nearby
the beat nodes) and $T$. Although the data do not allow a clear choice
between the two, the value $T_{N}=4$\,K seems to be too small for $x=0.04$.

At the same time, the experimental results cannot be described at $N\beta
^{\prime }<0.7$\,eV. The measured rate of shift is nearly twice as large as
that calculated at $N\beta ^{\prime }=0.6$\,eV
and $N\alpha ^{\prime }=-0.4$\,eV (the values given
in Ref.\,\cite{Furd88}) even if $T_{N}=0$ is
chosen. Although the exchange effects in studied systems with a strong
interband mixing are suppressed by SO splitting, such a discrepancy is
beyond the limits of experimental error. It is easy to verify that the
experimental data (energy position of LL's and its temperature shift)
presented in Refs.\,\cite{Mink} and \cite{Mink2} for bulk HgMnTe
with small $|E_{g}|$ also cannot be explained at $N\beta ^{\prime }<1,0\div
1.2$\,eV even for $T_{N}=0$. As already noted, the value of $N\beta ^{\prime
} $ reported in works on gapless HgMnTe falls typically within
$0.9\div 1.6$\,eV.

\subsection{Dingle temperature and scattering}

At calculations we suppose that $T_{D}$ for both spin-orbit branches is the
same, as it occurs for light and heavy holes in the bulk of a semiconductor.
This assumption has supporting experimental evidence. When three or more
beat nodes are observed in the oscillations, the ``partial'' oscillations
relating to different spin branches can be extracted from experimental $C(B)$
traces, using Fourier filtration and inverse Fourier transform. The $T_{D}$
values determined from fitting turn out to be close for both branches within
the accuracy of the analysis. At the same time, the amplitudes corresponding
to these branches can differ considerably (up to several times). Such a
difference is not surprising. It is clear that both the DOS and cyclotron
energy are different for two branches of a spectrum having significantly
different dispersions. As a result, the ``partial'' capacitance oscillations
for these branches differ not only by the period (which leads to the beat of
oscillations) but also by the amplitudes, even if the relaxation times are
equal. Although DOS at $B=0$ in a low-energy branch is higher, the
corresponding amplitudes can be less, because the lower cyclotron energy in
this branch leads to a lesser amplitude factor in Eqs. (\ref{oscgen}) and (%
\ref{oscapr}).

The relation between amplitudes depends on the subband occupancies (via the
effective mass), magnetic field, temperature and broadening parameters. For
different subbands, $B$ and $N_{s}$, the ratio of amplitudes $%
A_{c}^{-}/A_{c}^{+}$ can be below as well as above unity. However, the value
of $A_{c}^{-}/A_{c}^{+}$ decreases rapidly with increasing $N_{i}$. Such
behavior correlates well with decreasing ratio of Fourier line intensities $%
I_{i}^{-}/I_{i}^{+}$ experimentally observed (see Fig.\,\ref{fig4}). Thus the
difference in the amplitudes for different spin components of oscillations
mentioned in Refs.\,\cite{Rad85} and \cite{Luo} is to be expected
for 2D systems with strong SO interaction (without invoking spin-dependent
scattering).

The Dingle temperature $T_{D}$, determined from the fitting, slightly
increases (from $8\div 9$ to $13\div 15$\,K for ground subband) with the
increasing $N_{s}$. Such behavior is inherent in surface roughness
scattering \cite{Ando82,Rad86}. The $T_{D}(N_{s})$ dependencies are
essentially sublinear. This testifies that the efficiency of scattering is
suppressed with the increasing of the Fermi wave vector. This is possible if
the correlation length $\Lambda $ is large enough. The best agreement
between experimental and calculated values of $T_{D}$ is achieved at $%
\Lambda \approx (110\div 120)$\,\AA  and at the average interface
displacement $\Delta \approx (20\div 25)$\,\AA. Note that the screening
effects contribute significantly to the scattering, because in surface
layers on narrow-gap semiconductors, the Fermi wavelength turns out to be of
the same order of magnitude as 2D Thomas-Fermi screening length and $z$-size
of wave function. The above values of $\Lambda $ and $\Delta $ are many
times larger than tuose found for silicon \cite{Ando82} and substantially
exceed the corresponding values in III-V semiconductors also \cite{Rad86}.
This suggests much more disorder in the interface between ternary compounds
and their oxides. Using the $T_{D}$ values, the electron mobility can be
estimated as $0.8\times 10^{4}$\,cm$^{2}$/Vs in $i=0$ subband and $1.5\times
10^{4}$\,cm$^{2}$/Vs in $i=1$ subband for sample $S1$
at $N_{s}\sim 10^{12}$\,cm$^{-2}$ that is close to
a value $1\times 10^{4}$\,cm$^{2}$/Vs measured for
grain boundaries in $p$-HgMnTe with $x=0.1$ \cite{Grab84a}.

Somewhat larger values of $T_{D}$ are detected at small surface densities $%
N_{s}<5\times 10^{11}$\,cm$^{-2}$. The Coulomb scattering from the chargers
in oxide cannot cause this, because the theoretical calculations give the
values of relaxation times, which are larger by at least two orders of
magnitude. This conclusion has direct experimental evidence. It can be seen
in Fig.\,\ref{fig1}, that the charge localized in oxide differs by a factor
of several times for different sweep cycles. If the Coulomb scattering were
important, the amplitudes of oscillations corresponding to different cycles
(different $V_{fb}$) but to the same $N_{s}$ (to the same LL's number at
fixed magnetic field) would be different. Nevertheless, the oscillation
amplitudes are practically the same. A possible cause for the increase of
LL's broadening at small $N_{i}$ is intersubband scattering \cite{Rad86}.
\bigskip

{\bf Acknowledgements.}
This work was supported in part by the project Esprit N28890 NTCONGS EC
(Euro Community) and by the Grant from the Education Committee of Russian
Federation.

\end{document}